\documentclass[aps,twocolumn,showpacs,nofootinbib]{revtex4}
\usepackage{graphicx}
\newcommand{\beq}{\begin{equation}}
\newcommand{\eeq}{\end{equation}}
\newcommand{\beqa}{\begin{eqnarray}}
\newcommand{\eeqa}{\end{eqnarray}}

\def\opone{\leavevmode\hbox{\small1\normalsize\kern-.33em1}}

\newtheorem{theorem}{Theorem}

\begin{document}

\title{How much measurement independence is needed in order to demonstrate nonlocality?}

\author{Jonathan Barrett}
\affiliation{Department of Mathematics, Royal Holloway, University of London, Egham, Surrey TW20 0EX, United Kingdom}
\author{Nicolas Gisin}
\affiliation{Group of Applied Physics, University of Geneva, 1211 Geneva 4, Switzerland}

\date{\small\today}

\begin{abstract}
If nonlocality is to be inferred from a violation of Bell's inequality, an important assumption is that the measurement settings are freely chosen by the observers, or alternatively, that they are random and uncorrelated with the hypothetical local variables. We study the case where this assumption is weakened, so that measurement settings and local variables are at least partially correlated. As we show, there is a connection between this type of model and models which reproduce nonlocal correlations by allowing classical communication between the distant parties, and a connection with models that exploit the detection loophole. We show that even if Bob's choices are completely independent, all correlations obtained from projective measurements on a singlet can be reproduced, with the correlation (measured by mutual information) between Alice's choice and local variables less than or equal to a single bit.
\end{abstract}

\maketitle


Quantum nonlocality, whereby particles appear to influence one another instantaneously even though they are widely separated in space, is one of the most remarkable phenomena known to modern science \cite{EPR,BellSpeakable, Gilder, Maudlin}. Historically, this peculiar prediction of quantum theory triggered many debates and even doubts about its validity. Today, it is a well established experimental fact \cite{Aspect}.

The profound implications of quantum nonlocality for our world view remain controversial. But it is no longer considered as suspicious, or of marginal interest. It is central to our understanding of quantum physics, and in particular it is essential for the powerful applications of quantum information technologies. In 1991, A.~Ekert showed how shared entanglement could enable distant partners to establish a shared cryptographic key \cite{Ekert91}. Ekert's intuition is quite simple -- if there are no local variables, then no adversary could possibly hold a copy of these variables -- yet it came decades after the birth of quantum mechanics. In 2005, Barrett and co-workers used this intuition to show how, with no further assumptions about quantum theory, nonlocal correlations alone could ensure security of a secret key \cite{BarrettKentetal05}. Ac\'{i}n and co-workers extended this result, showing how to generate secret keys using simple nonlocal correlations that violate the well-known CHSH inequality \cite{Acinetal06}. Simultaneously, it was realized that Bell inequalities are the only entanglement witnesses that can be trusted in cases where the dimensions of the relevant Hilbert spaces are unknown \cite{Acinetal06}.

In an experimental demonstration of quantum nonlocality, measurements are performed on separated quantum systems in an entangled state, and it is shown that the measurement outcomes are correlated in a manner that cannot be accounted for by local variables. In order to conclude that nonlocality is exhibited, it is crucial for the analysis that the choices of which measurement to perform are freely made by the experimenters. Alternatively, they must be random and uncorrelated with the hypothetical local variables. It is well known that if the measurement settings are not random, but are in fact determined by the local variables, then arbitrary correlations can be reproduced \cite{Brans88}.

Here we reverse the argument. Taking for granted that quantum nonlocal correlations are produced, how independent must the measurement settings be assumed to be in order to rule out an explanation in terms of local variables \cite{Pawlowski}? We show that all correlations obtained from projective measurements on a singlet state can be reproduced with the mutual information between local variables and the measurement setting on one side less than or equal to one bit. This is not only of fundamental interest. If a cryptographic protocol is relying on the nonlocality of correlations for security, it is vital that there is no underlying local mechanism that an eavesdropper may be exploiting. The result shows that random number generators used to determine settings must be assumed to have a very high degree of independence from the particle source.

{\bf Bell Experiments.}
In an experimental test of a Bell inequality, two experimenters -- traditionally named Alice and Bob -- are spatially separated. They each control a quantum system, and the joint state of these two systems is an entangled state. Alice and Bob each perform a measurement on their quantum system, in such a way that the measurements are spacelike separated. Alice's measurement is chosen from a finite set of possibilities, as is Bob's. Call these measurement settings the {\it inputs} and label them $x$ for Alice's choice and $y$ for Bob's choice. The outcome of Alice's measurement (her \emph{output}) is labeled $a$ and Bob's $b$. By repeating this procedure many times and collating the data, Alice and Bob can estimate conditional probabilities of the form $P(a,b|x,y)$.

The question is then, when can the correlations produced in an experiment like this be explained by underlying local variables? Locality here implies that the outcome of Alice's measurement cannot be directly influenced by Bob's choice of which measurement to perform, and vice versa. More precisely, let the hypothetical underlying variables be 
denoted $\lambda$, and assume a distribution $P(\lambda)$. Then \emph{Bell locality} is the condition that
\begin{equation}\label{belllocality}
P(a,b|x,y,\lambda ) = P(a|x,\lambda ) \cdot P(b|y,\lambda ).
\end{equation}
If the correlations could in principle be explained as arising from underlying local variables, then they can be written in the form
\begin{equation}\label{localcorrelations}
P(a,b|x,y) = \sum_{\lambda} P(\lambda ) \cdot P(a|x,\lambda ) \cdot P(b|y,\lambda ).
\end{equation}
(Here and throughout, sums should be replaced with integrals when the variable is not discrete.)
The correlations cannot be written in this form if and only if they violate a Bell inequality.

Suppose that correlations are obtained which do violate a Bell inequality. What assumptions are necessary to conclude that this is indicative of nonlocality? There is a large literature on this topic, and a reader could do a lot worse than refer to Bell's original papers \cite{BellSpeakable}. But it is uncontroversial that the standard argument needs to assume that the inputs $x,y$ are freely, or at least independently and randomly chosen by Alice and Bob.
\begin{figure}
\includegraphics[width=9cm]{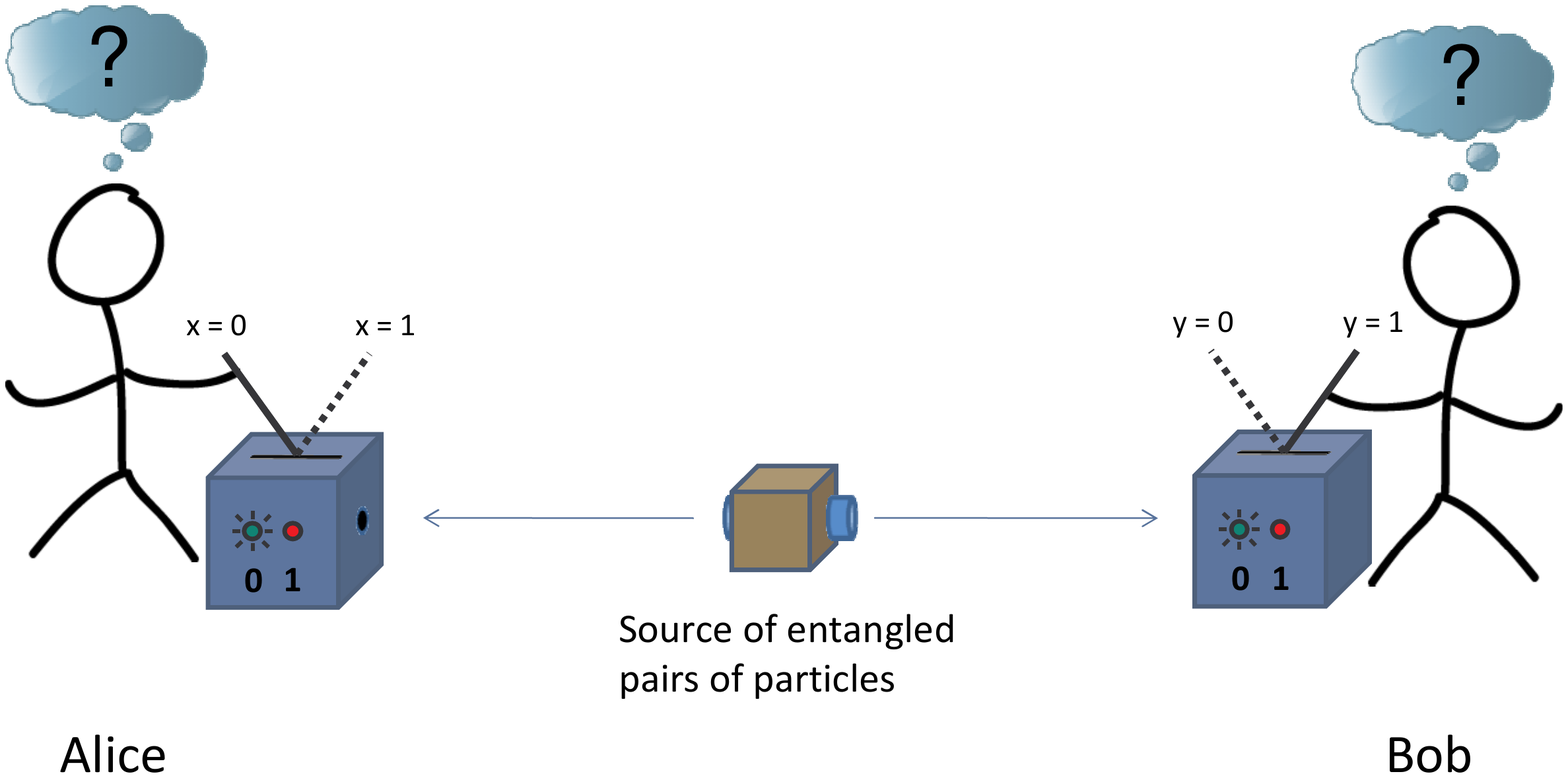}
\caption{\it Schematic illustration of an experiment for testing quantum nonlocality.}
\end{figure}


{\bf Non-independent measurement choices.}
We wish to analyze the case in which measurement settings can be correlated with local variables. Hence in addition to a set of correlations $P(a,b|x,y)$, assume a distribution over inputs $P(x,y)$. A \emph{correlated-settings} model is defined by a variable $\lambda$, and an overall joint probability distribution $P_{CS}(a,b,x,y,\lambda)$. The correlated-settings model \emph{reproduces} the given correlations and input distribution if $P_{CS}(a,b|x,y) = P(a,b|x,y)$ and $P_{CS}(x,y) = P(x,y)$.
The model is Bell-local if the distribution $P_{CS}$ also satisfies an equation of the form of Eq~(\ref{belllocality}). In this case
\begin{eqnarray}
&& P_{CS}(a,b|x,y) = \sum_{\lambda} P_{CS}(\lambda|x,y)\cdot P_{CS}(a,b|x,y,\lambda) \nonumber\\
&=& \sum_{\lambda} P_{CS}(\lambda|x,y)\cdot P_{CS}(a|x,\lambda)\cdot P_{CS}(b|y,\lambda).\label{localcorrelatedsettings}
\end{eqnarray}
Note the similarity with Eq.~(\ref{localcorrelations}). The usual kind of model, in which Alice's and Bob's inputs are assumed independent of the local variables, is a special case with $P_{CS}(x,y|\lambda) = P_{CS}(x,y)$. By Bayes' rule, this also gives $P_{CS}(\lambda|x,y) = P_{CS}(\lambda)$, and Eq.~(\ref{localcorrelatedsettings}) reduces to an equation of the form of Eq.~(\ref{localcorrelations}).

Another special case is the extreme case where the inputs are completely determined by $\lambda$, so that $x = f(\lambda)$ and $y = g(\lambda)$ for functions $f$ and $g$. Here, for all $x,y,\lambda$, $P_{CS}(x,y|\lambda) = 0$ or $1$. Any correlations can be produced by a Bell-local model of this form \cite{Brans88}.

In other models, $x$ and $y$ will be neither independent nor determined by $\lambda$. In general there are many numerical measures of the degree to which they are correlated. One particularly natural and simple measure is the \emph{mutual information}. Thus a correlated-settings model can be characterized by the mutual information between the measurement choices $x$, $y$ and $\lambda$:
\begin{equation}
I(x,y:\lambda) = H(x,y) + H(\lambda) - H(x,y,\lambda),
\end{equation}
where $H$ is the Shannon entropy. When $x$ and $y$ are independent of $\lambda$, $I(x,y:\lambda)=0$. If $x$ and $y$ are functions of $\lambda$ on the other hand, then $I(x,y:\lambda) = H(x,y)$. Since in general $I(x,y:\lambda) \leq H(x,y)$, this means that the mutual information is \emph{maximal} (with respect to a fixed distribution over inputs $P(x,y)$). 


In the context of correlated-settings models, the question is no longer whether quantum correlations violate a Bell inequality. The question is, how large must $I(x,y:\lambda)$ be if the correlations are to be reproduced within a Bell-local model? Alternatively, for a fixed degree of correlation $I(x,y:\lambda)$, do quantum correlations violate a Bell inequality by a sufficiently large margin that the model cannot possibly be Bell-local? This depends on the precise experiment. For example, if the input alphabet consists of only two choices, as with the CHSH inequality, then any correlations can be reproduced by a correlated-settings model with $I(x,y:\lambda)\leq 1$. But the question remains open for larger input alphabets. Intuitively one may think that for large alphabets, if Alice's choice is only mildly correlated with the variable $\lambda$, then a sufficient violation of a suitable Bell inequality should still rule out locality. But we shall see that this intuition is wrong, at least for maximally entangled qubits.



{\bf A connection to communication cost.}
One way to simulate quantum correlations, including nonlocal correlations, is for Alice and Bob to communicate with one another after inputs are chosen, but before outputs are produced. Any correlations can be produced this way if Alice and Bob share random data, and Alice simply informs Bob of her input choice. Hence, if Alice is choosing from $n_A$ possible inputs, then an obvious upper bound on how many classical bits need to be communicated is given by $\log n_A$. The minimum number of bits that must be communicated, on the other hand, provides a natural way of quantifying the amount of nonlocality inherent in a given set of correlations. This interesting line of research was started by T.~Maudlin \cite{Maudlin} and independently by Tapp et al. and by Steiner \cite{TappCB99,Steiner00}. It culminated with a model of Toner and Bacon, who proved that one single bit of communication from Alice to Bob is sufficient to simulate all correlations obtained from arbitrary projective (i.e. Von Neumann) measurements on two qubits in a maximally entangled state \cite{TonerB03}. It was extended to all two-party correlations, though ignoring the marginals, in \cite{RegevToner}.

A slightly more precise description of a communication model for simulating quantum correlations is as follows. The random data shared between Alice and Bob is a variable $\mu$, with distribution $P_{C}(\mu)$. Assuming that Alice communicates first, Alice sends a bit string $c_1$ to Bob. The string $c_1$ can depend on $\mu$ and on Alice's input $x$. It may also depend on further random data held by Alice, but this data can without loss of generality be absorbed into $\mu$. Hence assume that $c_1 = c_1(\mu,x)$. Depending on the protocol Bob may now send a bit string $c_2$ to Alice, where again without loss of generality, we assume that $c_2$ is a function of $\mu$, $c_1$ and $y$. This continues for a total of $k$ messages. The entire conversation between Alice and Bob on a particular round is the sequence $m = c_1, c_2, \ldots, c_k$, and is a function $m = f(x,y,\mu)$ of $x$, $y$, and $\mu$. Finally, Alice outputs $a$, where again absorbing all randomness into $\mu$, $a$ can be assumed to be a function $a = g(x,\mu,m)$ of her input, the shared random data and the conversation. Similarly, Bob outputs $b$, which is a function $b=h(y,\mu,m)$. The protocol then terminates.

For each pair of inputs, a communication model defines a distribution $P_{C}(a,b,m,\mu|x,y)$, and reproduces correlations $P(a,b|x,y)$ if $P_{C}(a,b|x,y)=P(a,b|x,y)$. If a distribution over inputs $P(x,y)$ is also given, the conversation $m$ has a well-defined distribution $P(m) = \sum_{x,y} P_{C}(m|x,y)\cdot P(x,y)$. Let the entropy of this distribution be $H(m)$.


The connection with correlated-settings models is expressed in the following
\begin{theorem}
Consider a fixed input distribution $P(x,y)$, and a communication model with total conversation $m$. Then there exists a Bell-local correlated-settings model, which reproduces the same correlations and $P(x,y)$, with $I(x,y:\lambda) \leq H(m)$.
\end{theorem}
{\bf Proof.}
Given $P(x,y)$ and a communication model as above, construct a correlated-settings model as follows. Define $\lambda$ as the pair $\lambda = (\mu, m)$. In the communication model, $m$ is the conversation between Alice and Bob, but in the correlated-settings model there is no communication, and $\lambda$ represents the underlying variable. Define the correlated-settings model by $P_{CS}(a,b,x,y,\lambda) = P(x,y) P_{C}(a,b,\mu,m|x,y)$. It is easy to see that the resulting model reproduces $P(x,y)$ and the same correlations as the communication model. Bell locality is satisfied since
\begin{eqnarray*}
P_{CS}(a,b|x,y,\lambda) &=& P_{C}(a,b|x,y,\mu,m) \\
&=& P_{C}(a|x,\mu,m)\cdot P_{C}(b|y,\mu,m)\\
&=& P_{CS}(a|x,\lambda) \cdot P_{CS}(b|y,\lambda).
\end{eqnarray*}
The mutual information between $\lambda$ and $(x,y)$ is
\beqa
I(\lambda:x,y) &=& I(\mu:x,y) + I(m:x,y|\mu)\\
&=& 0 + H(m|\mu)-H(m|\mu,x,y)\\
&=& H(m|\mu)\\
&\le& H(m).
\eeqa
The first line uses the chain rule \cite{InfoTh}, the second line follows from the fact that $\mu$ is assumed to be independent of the inputs $x,y$, and the third line holds because $m$ is a function of $\mu$ and $x,y$. This concludes the proof.

Intuitively, $\lambda = (\mu,m)$ restricts Alice's and Bob's choices to inputs $(x,y)$ such that $m = f(x,y,\mu)$.
We stress the generality of this result: the communication model may have involved two way communication, and may have been such that the communication on a given run is unbounded. As long as $H(m)$ is finite, the bound is useful.

We now apply the theorem with reference to the Toner-Bacon communication model. Here, $x$ and $y$ are arbitrary projective measurements. The shared random data $\mu$ takes the form of two vectors on the Bloch sphere and the distribution $P(\mu)$ is such that the two vectors are independent, with each uniformly distributed. For our purposes most of the details of the model do not matter, and we refer the reader to \cite{TonerB03} for a description and proof that it reproduces quantum correlations for projective measurements on a singlet. It suffices to know that each round, the conversation $m$ is a single bit communicated from Alice to Bob. The bit $m = f(x,\mu)$ is a function of Alice's input $x$ and $\mu$.

The resulting correlated-settings model also reproduces all correlations from projective measurements on a singlet. In general, the actual value of $I(x,y:\lambda)$ depends on the distribution over inputs $P(x,y)$. The theorem tells us that for any such distribution, $I(x,y:\lambda)\leq H(m) \leq 1$, since $m$ is a single bit. In particular, if Alice's and Bob's inputs are independent, with each chosen from a uniform distribution over all possible directions, it is easy to verify that $I(x,y:\lambda) \approx 0.85$.

In a typical Bell-type experiment with a singlet, Alice and Bob will be choosing from finite sets of measurements. The Toner-Bacon model can still be applied, with the distribution $P(x,y)$ having support only on these finite sets. In this case too, $I(x,y:\lambda)\leq 1$. The Bell-local correlated-settings model reproduces the quantum correlations no matter how large the input alphabets, and no matter what Bell inequality is being tested.

Finally, in the Toner-Bacon model, Bob's setting is of course independent from the communication he receives from Alice. It follows that in the derived correlated-settings model, Bob's setting is independent from $\lambda$, i.e., $I(y:\lambda)=0$.



{\bf The detection loophole.}
In a real Bell experiment, detection is inefficient. A typical analysis of the experiment estimates the correlations $P(a,b|x,y)$ from those runs on which both Alice's and Bob's detectors clicked, and simply ignores all the other runs. Such an analysis is valid, as long as it is assumed that the probability of a detector clicking is independent of the hypothetical local variables. This is sometimes called the \emph{fair sampling assumption}.

It is possible to reproduce nonlocal correlations with a model that is Bell-local, but which violates the fair sampling assumption \cite{PearleDetLoophole,SantosDetLoophole}. Denote the event that Alice's (Bob's) detector clicks by $D_A$ ($D_B$). A \emph{detection-efficiency} model is defined by a variable $\lambda$, with distribution $P_{DE}(\lambda)$, independent of $x,y$, and for each $x,y$ a distribution $P_{DE}(a,b,D_A,D_B|x,y,\lambda)$. Bell-locality is the condition that $P_{DE}(a,b,D_A,D_B|x,y,\lambda) = P_{DE}(a,D_A|x,\lambda)\cdot P_{DE}(b,D_B|y,\lambda)$. The model reproduces correlations $P(a,b|x,y)$ if
\begin{equation}\label{deteffcorrelations}
P(a,b|x,y) = P_{DE}(a,b|x,y,D_A,D_B).
\end{equation}

The efficiencies of Alice's detectors in this model are given by $P_{DE}(D_A|x) = \sum_{\lambda} P_{DE}(\lambda) P_{DE}(D_A|x,\lambda)$, and similarly for Bob's. If the detection efficiencies in a real experiment are high enough, and if a Bell inequality is violated by a large enough margin, then a Bell-local detection-efficiency model reproducing the correlations can be ruled out.

Given a distribution $P(x,y)$, and a detection-efficiency model reproducing correlations $P(a,b|x,y)$, it is easy to construct a correlated-settings model which also reproduces $P(x,y)$ and $P(a,b|x,y)$. Simply define
\begin{equation}
P_{CS}(a,b,x,y,\lambda ) = P(x,y)\cdot P_{DE}(a,b,\lambda | x,y,D_A,D_B).
\end{equation}
It is easy to show, using Eq.~(\ref{deteffcorrelations}), that the correlated-settings model reproduces $P(x,y)$ and $P(a,b|x,y)$.


The construction can be applied, for example, to the Gisin-\&-Gisin detection-efficiency model \cite{GisinG99}, which reproduces correlations from arbitrary projective measurements on a singlet. The efficiencies are independent of $x$ and $y$ and are given by $P(D_A) = 1/2$, $P(D_B) = 1$. In this model $\lambda$ is a vector on the Bloch sphere, and $P_{DE}(\lambda|x,y,D_A,D_B)=|\lambda.\vec x|/2\pi$ where $\vec x$ denotes the Bloch vector representing the measurement $x$. In the resulting correlated-settings model, $I(x,y:\lambda)$ depends on $P(x,y)$. If $x$ and $y$ are independently and uniformly distributed, it is easy to show that $I(x,y:\lambda) = I(x:\lambda) \approx 0.28$. This is an even lower value than was obtained above using the Toner-Bacon model.

Finally, note that in the case of communication models, there was a connection between $I(x,y:\lambda)$ and the amount of communication (as measured by the entropy of the conversation's distribution). With detection efficiency models, we expect there to be a connection between $I(x,y:\lambda)$ and the detection efficiencies. Exploring this connection is an interesting direction for future work.



{\bf Conclusion.}
It is well known that in order to derive nonlocality from violation of a Bell inequality, one has to assume that there is no correlation between the hypothetical local variables $\lambda$ and the experimenters' measurement choices $x$ and $y$. We have investigated how much correlation could be allowed if quantum predictions are to remain incompatible with local variables. Surprisingly, no matter how large the alphabet size of the inputs, correlations from projective measurements on a singlet can be reproduced, with the mutual information between Alice's input and local variables not more than one bit, and Bob's input completely independent. This deserves further investigation. Future work should analyze more general scenarios, involving generalized measurements, partial entanglement \cite{partialEntanglement}, higher dimensional Hilbert spaces and more than two parties. It would also be instructive to consider measures of correlation other than the mutual information.

Finally, let us emphasize the change in paradigm since the old EPR paper \cite{EPR}. If, contrary to EPR, one accepts nonlocality as a fact, then not only can one develop powerful applications in quantum information science, like device-independent quantum key distribution \cite{BarrettKentetal05, Acinetal06}, but moreover one can upper bound the lack of free choice of the players !

\small

\section*{Acknowledgment}
JB is supported by an EPSRC Career Acceleration Fellowship. NG acknowledges support from the ERC advanced grant QORE and the Swiss NCCR-QP. After completion of this work M.~J.~W.~Hall posted a related interesting paper: arXiv:1007.5518 (PRL 105, 250404 (2010)). We thank him and C.~Branciard for constructive discussions.


\begin{thebibliography}{99}
\bibitem{BellSpeakable} J.~S.~Bell, {\it Speakable and Unspeakable in Quantum Mechanics: Collected papers on quantum philosophy}, Cambridge University Press, Cambridge, 1987, revised edition 2004.

\bibitem{Maudlin} T.~Maudlin, {\it Quantum Non-Locality and Relativity}, Blackwell Publishers, 2nd edition, 2002.

\bibitem{Gilder} L.~Gilder, {\it The Age of Entanglement}, Alfred~A.~Knopf, 2008.

\bibitem{EPR} A.~Einstein, B.~Podolsky and N.~Rosen, Phys. Rev. {\bf 47}, 777 (1935).

\bibitem{Aspect} A.~Aspect, Nature \textbf{398}, 189 (1999).

\bibitem{Pawlowski} see also: J.~Kofler et al., Phys. Rev. A {\bf73}, 022104 (2006); M.~Pawlowski et al., arXiv:0903.5042.

\bibitem{2ndQrevolution} A.~Aspect, Nature {\bf466}, 866 (2007).


\bibitem{Ekert91} A.~K.~Ekert, Phys. Rev. Lett. {\bf 67} 661 (1991).

\bibitem{BarrettKentetal05} J.~Barrett, L.~Hardy and A.~Kent, {\it Phys. Rev. Lett.} {\bf95},
010503 (2005).

\bibitem{Acinetal06} A.~Acin, N.~Gisin and Ll.~Masanes, {\it Phys. Rev. Lett.} {\bf97}, 120405
(2006).



\bibitem{Brans88} C.~Brans, Int. J. Theoret. Phys. {\bf27}, 219 (1988)

\bibitem{TappCB99} G.~Brassard, R.~Cleve and A.~Tapp, Phys. Rev. Lett.
{\bf 83}, 1874 (1999).

\bibitem{Steiner00} M.~Steiner, Phys. Lett. A {\bf 270}, 239 (2000).

\bibitem{TonerB03} B.~F.~Toner and D.~Bacon, Phys. Rev. Lett. {\bf 91}, 187904 (2003).

\bibitem{RegevToner} O.~Regev and B.~F.~Toner, Proceedings of 48th Annual IEEE Symposium on Foundations of Computer Science (FOCS 2007).

\bibitem{InfoTh} T.~M.~Cover and J.~A.~Thomas, {\it Elements of information theory}, Wiley, New York, 1991.

\bibitem{PearleDetLoophole} P.~Pearle, Phys. Rev. D {\bf2}, 1418 (1970).

\bibitem{SantosDetLoophole} E.~Santos, Phys. Rev. A {\bf46}, 3646 (1992).

\bibitem{GisinG99} B.~Gisin, N.~Gisin, Phys. Lett. A {\bf 260}, 323 (1999).

\bibitem{partialEntanglement} For two partially entangled qubits, there is a 2 bits communication model \cite{TonerB03}; surprisingly, however, it is still unknown whether this is optimal or whether a single bit would suffice.


\end{thebibliography}
\end{document}